\documentclass{article}
\usepackage{ijcai19}

\usepackage{times}
\usepackage{soul}
\usepackage{url}
\usepackage[hidelinks]{hyperref}
\usepackage[utf8]{inputenc}
\usepackage[small]{caption}
\usepackage{graphicx}
\usepackage{subcaption}

\usepackage{amsmath}
\usepackage{booktabs}
\usepackage{algorithm}
\usepackage{algorithmic}

\usepackage{rotating}
\usepackage{placeins}

\usepackage{lipsum}
\usepackage{multicol}

\title{Neural Networks for Predicting Human Interactions in Repeated Games\thanks{The supplementary materials for this paper are available at: \url{http://www.cs.huji.ac.il/\~gali_n/supp.pdf}}}

\author{
Yoav Kolumbus$^{1,2}$
\And
Gali Noti$^{1,3}$
\affiliations
$^1$The School of Computer Science and Engineering, Hebrew University of Jerusalem, Israel.\\
$^2$Racah Institute of Physics, Hebrew University of Jerusalem, Israel.\\
$^3$Federmann Center for the Study of Rationality, Hebrew University of Jerusalem, Israel.\\
\emails
yoav.kolumbus@mail.huji.ac.il,
gali.noti@mail.huji.ac.il 
}

\begin{document}

\maketitle

\begin{abstract}
We consider the problem of predicting human players' actions in repeated strategic interactions. Our goal is to predict the dynamic step-by-step behavior of individual players in previously unseen games. We study the ability of neural networks to perform such predictions and the information that they require. We show on a dataset of normal-form games from experiments with human participants that standard neural networks are able to learn functions that provide more accurate predictions of the players' actions than established models from behavioral economics. The networks outperform the other models in terms of prediction accuracy and cross-entropy, and yield higher economic value. We show that if the available input is only of a short sequence of play, economic information about the game is important for predicting behavior of human agents. However, interestingly, we find that when the networks are trained with long enough sequences of history of play, action-based networks do well and additional economic details about the game do not improve their performance, indicating that the sequence of actions encode sufficient information for the success in the prediction task. 
\end{abstract}

\vspace{-10pt}

\section{Introduction} \label{intro}
\vspace{-2pt}

Predicting human agents' decisions is 
one of the most 
attractive goals and the subject of extensive study both in the academy and in the industry (e.g., \cite{rosenfeld2018,erev2013}). 
Game theory provides a framework for studying players' decisions in strategic interactions (``games''), 
under the assumption that players are rational, i.e., the assumption that players choose their actions so as to maximize their utility. 
Research in {\em behavioral} game theory asks what happens when the players are humans, who are not completely rational or just not able to optimize perfectly (
e.g., \cite{camerer2011,KL12,Kagel1995}). 
In this paper we focus on predicting human players' actions in a setting where a game is played {\em repeatedly}. 
In the repeated game setting the dynamics allow for 
greater complexity
and 
game theory predicts the possibility of multiple equilibria, but, on the other hand, there may be behavioral patterns 
that can be utilized for the prediction.

The prediction task can be addressed 
at different levels of aggregation 
and generalizability. 
To demonstrate the differences, consider two players who repeatedly play the 
``matching pennies'' game. 
At each repetition of the game, each of the players 
turns one penny to either heads or tails, and if both coins match, player 1 receives the two pennies, or otherwise player 2 receives them. 

One approach is to predict the average aggregate behavior of a population of players 
in the game. 
Works that provide such a static prediction 
showed 
that they succeed to generalize across 
games, i.e., they are able to learn model parameters in a given set of games and then predict the average behavior 
of players 
in a new game. 
This category of predictors includes equilibrium models 
that predict stationary distributions of the players' actions, such as Nash equilibrium 
or other equilibrium concepts that are based on behavioral modeling (e.g., \cite{qre,ibe}).   
At this level 
of the prediction task, i.e., the aggregate prediction, 
the best performance one can hope for is to correctly predict the actual empirical distribution. 
However, 
such a prediction 
is not really satisfying 
in a context of 
repeated games; returning to our matching pennies example, a player who uses an aggregate predictor, even the best one, would not be able to predict better than the uniform distribution played by a population of players, and thus would not be able to utilize this predictor to ``win'' the game. 

A second possibility is to specialize in the specific game by learning the tendencies of players in the game, and then trying to predict the actual actions of players step-by-step. 
Then, it is possible to learn, for example, from many instances of the matching pennies game, that players would tend to, say, choose 
3 times heads and then 3 times tails. 
Such a predictor would allow a user to win the game, i.e., manage to achieve above-average winning rates if utilized for opponent modeling -- but at the cost of being tailored to the specific game.

In this study we address the combined challenge of predicting the dynamic {\em step-by-step} behavior of players, in {\em previously unseen} games.  
The behavioral economics literature suggests learning models that are natural candidates for this task. These models update their prediction at every step according to the previous result's feedback. They aim to model human learning processes and in principle should provide accurate predictions of human players' actions. Examples are the classic reinforcement learning model \cite{rl} and normalized fictitious play \cite{nfp,nfp2010}. 
However, these models usually update their predicted distributions in small steps, and either converge or move slowly between quasi-static distributions, and it is questionable whether 
they 
are suitable for 
predicting dynamic action patterns of individual players. 
For example, the above-mentioned models would fail to predict the actions of a player who periodically alternates between heads and tails.

We study the ability of neural networks to answer our prediction challenge and perform ``behavior modeling.''  
For the case where the game is played only once, \cite{NIPS2016} 
show that neural networks are able to outperform behavioral models in predicting human behavior. 
Unlike the behavioral learning models, neural networks learn in advance a {\em fixed} function from 
one 
set of instances 
and then use this same function to perform a prediction on 
new instances.
It is not clear a priori whether indeed behavior in repeated games 
can be described by a single fixed function that can be general to a wide collection of games, or whether 
behavior can be predicted only through 
online feedback. 
We note that 
our  
behavior-modeling prediction task 
resembles the 
task of language modeling, where the next word in a text is predicted 
from the previous words in the text 
(e.g., \cite{lang_modeling1,lang_modeling2}). From a learning perspective, 
the two can be viewed as similar sequence-prediction tasks, 
but 
in behavior modeling all sequences are in principle allowed, and we do not know in advance whether players' actions have any structure that can be generalized across 
games, and that may be informative for predicting future actions 
in 
new
games.

In our supervised learning framework, in the training stage the input to the network is the game history up to time $t$ 
and the output is a prediction of the player's action at the next step $t+1$. 
The model parameters are optimized according to a loss function with respect to the 
actions that were played in practice. 
We use for evaluation the 2x2 two-player game dataset of \cite{SC2008}, which consists of 12 games, each with a unique (fully) mixed Nash equilibrium, from experiments with human participants. 
Such games are commonly used in the behavioral literature as a natural testing ground for studying behavior in repeated interactions, and were shown 
to give rise to various behavioral biases (e.g., \cite{erev2007learning,KL12}). 
Each game was played in multiple independent sessions by human players over 200 game periods.  
To measure the generalizability of our models, for each game we train a separate model on the other 11 games and evaluate the model's performance on the 12th game. We take 
the overall performance of a model type to be 
the average performance over these 12 models.

We evaluate the performance of 
two classic network types---the Multi-Layer Perceptron (MLP) 
and Convolutional Neural Networks (CNN)\footnote{We also tried additional network architectures, including LSTM, inception and residual networks, but this did not improve the prediction results. See Section \ref{models} for more details.}---in comparison 
with established 
models from behavioral economics and with 
networks that are allowed to specialize in the specific game in their training stage. 
Our measures of success are the cross-entropy loss and the prediction accuracy per game period obtained by a model, as well as  
the economic value gained by using a model as an oracle of the opponent's behavior by a rational player. 
See Section \ref{task} for more details about 
the prediction task, and Section \ref{models} for more details about the models.

We start with the most basic input to the network, which consists only of the actions played by the two players thus far in the 
previous game periods. 
That is, as a first step in our analysis, we ignore any additional information about the payoffs of the game and any strategic considerations that the players may have, and observe, as in a black box, only what the players actually did. 
We find that in our dataset these actions are sufficient input 
for network models to predict future human behavior in previously unseen games better than 
the static-distribution predictors (Section \ref{static}) and the dynamic learning models (Section \ref{dynamic}), according to all measures: loss, accuracy, and economic value. 
We also compare these ``domain-adapted'' network models 
to ``specialized'' networks that were trained and tested on independent sessions of the same game.
Specialized models have the advantage of expert knowledge in a specific game, but on the other hand, focusing on a specific domain reduces the amount of available training data and creates a tradeoff on the model's performance.
In our dataset the specialized network models
performed worse 
than the domain-adapted network models, but still performed better than all the non-network models  
(see Section \ref{eval}). 

Next, we compare 
the action-based network models described above with econ-aware networks that in addition to the observed actions get as input 
economic information about the game. 
We find that when the networks are trained with short history sequences, the additional economic information is useful and improves the performance of the networks. 
However, interestingly, when the history sequence is long enough, the additional information does not improve the performance, 
indicating that the actions encode all the information necessary to perform the prediction at the accuracy level the networks obtained. 
See Section \ref{econ} for more details. 

To summarize, the main contributions of the present paper are as follows.
This paper is the first to study the task of predicting human step-by-step behavior in repeated 
normal-form 
games as an offline learning problem, and to show that it is possible to generalize the predictions across different games. 
More specifically, this paper is the first to study action prediction in repeated games with neural network models. 
We show that networks are suitable for the task and outperform established models from behavioral economics with a significant gap. 
In addition, we show for the first time that it is possible to predict future behavior based only on observed behavior, without knowing any economic details about the game. 
This observation may be useful in other games where the utilities are the private information of the players and do not appear in the data. 
From a behavioral modeling perspective, our results show that the deviations of human behavior from rational play have common characteristics that can be learned and transferred between different games via fixed functions and without explicit modeling of player preferences.
For a further discussion of our contributions, 
see Section \ref{conclusion}. 

\section{Setting and the Prediction Task} \label{task}

For the general repeated game setting, consider $n$ players repeatedly playing a game. 
Denote by $A_i$ the action space of player $i$ and by $a_i^t \in A_i$ the action played by player $i$ at period $t$.
Denote by $a_{-i}$ an action profile of the players except for player $i$ (i.e., $a_{-i}=(a_1,..., a_{i-1},a_{i+1}, ..., a_n)$).
The game is defined by utility functions $u_i(a_i, a_{-i})$ that determine for each player $i$ his payoff from choosing action $a_i$ when the other players play $a_{-i}$.

We consider the dataset of 12 different 2x2 games from the experiment of \cite{SC2008}.
In 
2x2 games 
there are two players---the row player and the column player---who are repeatedly playing according to a utility function that is defined by a fixed payoff matrix. 
At each game period, each player has a binary choice: the row player chooses an action in 
$\{Up, Down\}$ 
and the column player chooses an action in 
$\{Left, Right\}$. 
The payoff matrices of the 12 games 
are presented in 
the supplementary materials. 
Each of the 12 games 
has a unique mixed Nash equilibrium in which each player plays each of his actions with positive probability.

The data consist of 108 
independent experimental 
sessions: 12 sessions for each of games 1-6, and 6 sessions for each of games 7-12. 
In each session there were four row players and four column players, who repeatedly played a game over 200 periods. 
See the supplementary materials 
for more details on the experimental setup. 
We exclude from the data the first and last 10 periods of play of each game in both training and testing, to remove possible effects related specifically to the beginning or ending of a sequence of play. The overall size of the data is 155,520 actions for prediction. 

The behavior modeling task is to predict for player $i$ at each time $t=1...T$ the player's next action $a_i^{t+1}$.
The input 
is the history of actions that were actually played by the two players until time $t$, and possibly additional economic information such as the game payoff matrices or functions thereof. 
The output of the prediction model is a probability distribution over the actions $A_i$.  
At the training stage, the model observes sequences of actions played by human players in a set $G$ of games, on which the model optimizes its prediction, while the test is performed over sequences of play in game $g' \notin G$.

The metrics we consider are the cross-entropy loss, prediction accuracy and economic value.
More formally, let $y_i^t \in \left\{ 0,1 \right\}$ be the action of player $i$ at time $t=1...T$ (such that $0$ and $1$ stand 
for {\em Up} and {\em Down} when predicting for a row player, or {\em Left} and {\em Right} for a column player), and let $\hat{y}_i^t \in \left[ 0,1 \right]$ be the predicted probability of action $0$. The cross-entropy loss of a model in game $g$ with $n$ players is:
$\mathcal{L}_g =  - \frac{1}{n} \sum_i \frac{1}{T} \sum_t y_i^t \cdot ln(\hat{y}_i^t) + (1 - y_i^t) \cdot ln(1 - \hat{y}_i^t)$. 
The prediction accuracy metric is the percentage of accurate predictions when choosing the action that maximizes likelihood in respect to the predicted distribution, and can be written in a vector notation as 
$\frac{100}{nT} \sum_{i=1}^n \left[ \textbf{y}_i  \cdot (\textbf{1} - \hat{\textbf{y}}_i) + (\textbf{1} - \textbf{y}_i) \cdot \hat{\textbf{y}}_i \right]$, where \textbf{y} is the vector $(y_i^1,...,y_i^T)$ and $\textbf{1}$ has the same dimensions as $\textbf{y}$ with all entries set to unity. 

The economic value of a model quantifies how much a player could have earned had he rationally followed (i.e., optimized against-) the model predictions. 
It is 
the ``utility gained by this model,'' as percentages from the optimal utility.  
Specifically, the utility gained by a model is calculated by using this model 
as an oracle for a player regarding the distribution over the actions of the opponent at the next step. The player then best responds to this predicted distribution. 
I.e., the action of player $i$ at time $t$ given a prediction $(p_0, p_1)$ of the opponent's distribution of actions is
$a_i^t = arg\max_{{a} \in \left\{ 0,1 \right\}} \left( p_0 \cdot u_i(a,0) + p_1 \cdot u_i(a,1)\right)$. 
The economic value metric is counted according to the actual actions $a_{-i}^t$ played by the opponent at each time $t$, 
and 
is taken 
for each game $g$ with $n$ players over all decision steps of all players as percentages from the optimal utility: 
$100 \cdot  \frac{\sum_{i} \sum_{t} u_i(a_i^t, a_{-i}^t)}{\sum_{i} \sum_{t} u_i(opt_i^t, a_{-i}^t)}$, where $opt_i^t = arg\max_a u_i(a,a_{-i}^t)$ is the optimal choice in hindsight for player $i$ at time $t$.

After we calculate for each model the three metrics for each of the games as described above, 
the loss, accuracy, and economic value of a model type are defined as the average of each metric over 
the different games, weighted by the number of prediction steps in each game. 
So that the performance of a model type is the weighted average of performances of 12 models of this type, each tested on a game which it did not observe in the training data and was not optimized for.

\section{The Neural Network Models} \label{models}

We test and compare the performance of neural network models in the task of behavior modeling, as defined in Section \ref{task}. 
All neural network 
models were implemented in Keras 2.1.4 \cite{chollet2015keras} and Tensorflow 1.5.0 \cite{tensorflow2015-whitepaper}. 
The hyper parameters for the models 
and 
the number of layers and size of each layer were optimized on a 
validation dataset that 
consisted of $5\%$ of the training sequences. 
Each input to the model is a sequence of the $k$ previous steps. 
In principle $k$ could be the entire game history, but we use $k=20$ that was sufficient for our prediction task (see Section \ref{econ}). 
The output is a probability distribution over the actions $A_i$ of player $i$.

We 
focus on the performance of two classic network types: Multi-Layer Perceptrons (MLP) and Convolutional Neural Networks (CNN). The results we present in the next section show that even with basic architectures, 
network models outperform the behavioral models.
We also tested the performance of additional network architectures, however these experiments showed similar or inferior results to the standard CNN. 
Specifically, our experiments with   
inception networks \cite{inception} and residual networks \cite{residual} (where we replaced the standard convolutional layers of the CNN with inception modules or with residual blocks) yielded similar results to the standard CNN model. 
We also experimented with LSTM networks \cite{hochreiter1997long,lstm99}, which are often used for sequence prediction in other contexts (e.g., in language modeling \cite{sundermeyer2012lstm,greff2017lstm}). However, single-layer LSTM models with 100 recurrent units achieved inferior results when compared to the other network models, with similar performance to the inertia rule.\footnote{Specifically, the average accuracy, cross-entropy and economic value of the inception networks were: 79.5\%, 0.420 and 87.53\%, respectively; of the residual networks: 79.3\%, 0.423 and 87.41\%, respectively; and of the LSTM networks: 74.5\%, 0.523 and 82.50\%.}

\begin{figure*}[t] 
\begin{subfigure}{.33\textwidth} 
  \includegraphics[scale=0.37]{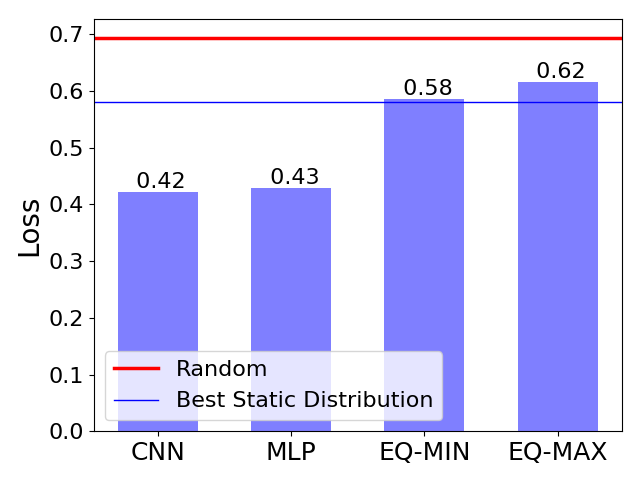} 
  \caption{\label{fig:bar_static_loss}}
\end{subfigure}
\begin{subfigure}{.33\textwidth}
  \includegraphics[scale=0.37]{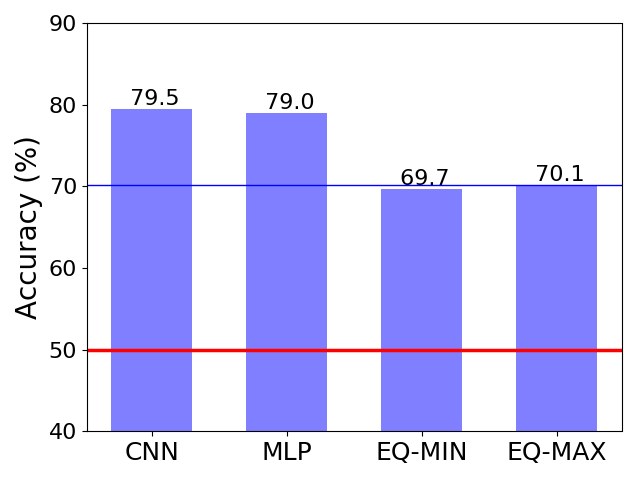} 
  \caption{\label{fig:bar_static_acc}}
\end{subfigure}
\begin{subfigure}{.33\textwidth}
  \includegraphics[scale=0.37]{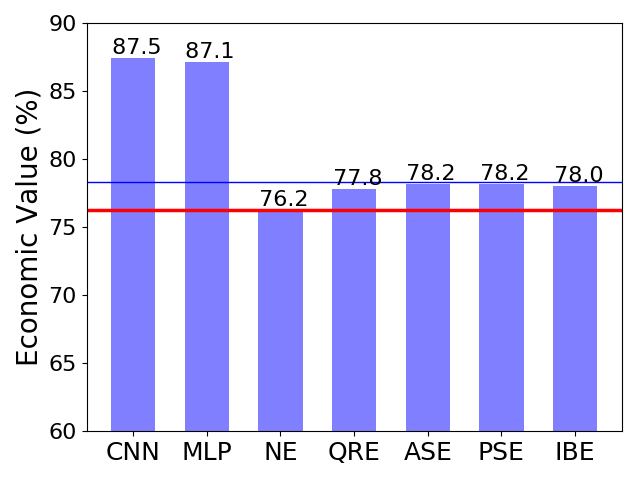}
  \caption{\label{fig:bar_static_value}}
\end{subfigure}
\caption{Comparison to static distributions: (\ref{fig:bar_static_loss}) cross-entropy loss, (\ref{fig:bar_static_acc}) prediction accuracy, and (\ref{fig:bar_static_value}) economic value of the network models and the equilibrium models (see Section \ref{static} for more details). The blue horizontal line indicates the performance of the best static distribution benchmark, and the red line indicates the performance of the random benchmark.
 \label{fig:bar_static}}
\end{figure*}

\subsection{Multi-Layer Perceptrons (MLP)}

MLPs are fully connected feed-forward neural networks, which provide a benchmark for the ability of a simple neural network architecture to model and predict human player's actions based on a learned training dataset. We use networks with two hidden layers, each with 512 units with a point-wise rectified linear activation function (ReLU), and a two-unit output layer with a softmax activation function to output a probability distribution over the player's actions. 
Training is performed with dropout regularization \cite{dropout} with a weight deletion rate of 0.3, and we use the Adam optimizer \cite{adam} with a learning rate of $0.0002$ and a batch size of 64 sequences.
These networks get the information regarding each input sequence as a single vector; i.e., they do not get the temporal dimension of the data explicitly as a separate dimension of their inputs. 
Experiments with larger networks with up to 5 layers and layer sizes of up to 1024 did not change the results. A network with a single hidden layer (and otherwise same details) did yield worse performance than the two-layer model.
  
\subsection{Convolutional Neural Networks (CNN)}

CNNs are often used for temporal signal processing (see, e.g., \cite{abdel2014convolutional,gehring2017convolutional}) and are natural candidates for our step-by-step sequence prediction task. CNNs are capable of representing local temporal relations while maintaining a small number of parameters. In the repeated game setting these can be temporally-local action and response patterns, which may appear at different places in the observed sequences. The player's and the opponent's actions are taken as input to the network in two separate channels, and the convolution is performed over time. We use a network with two convolution layers, each with 64 $5 \times 1$ filters, one fully connected layer with 256 units and ReLU activation, and a softmax output layer with two units. We use the same regularization and optimization methods as for the MLP networks. 
Experiments with larger networks, with up to 5 convolution layers and 2 fully connected layers, did not show any advantage and achieved similar results. We also tested two single-channel input CNN variants, one with the same number of parameters as our two-channel model, and one with a double number of parameters, and found that the two-channel model had a slight advantage over these variants.

\section{Evaluation Results} \label{eval}

In this section we evaluate the ability of the 
MLP and CNN networks
to predict humans' 
step-by-step behavior across games, as described in Section \ref{task}.
Here we provide the networks with the most basic input that consists only of the actions played by the two players in the preceding $k=20$ game periods (in Section \ref{econ} we show that this sequence length was sufficient).
That is, we ignore any information about the game payoff structure or strategic considerations of the players, and only observe what the players actually did. 
As we show next, these actions are sufficient for the network models to predict human future behavior in previously unseen games better than the other models we have tested. 

\begin{figure*}[t] 
\begin{subfigure}{.33\textwidth} 
  \includegraphics[scale=0.37]{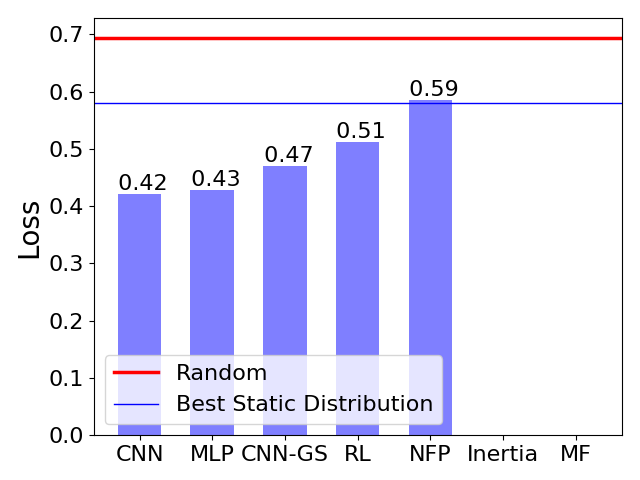} 
  \caption{\label{fig:bar_dynamic_loss}}
\end{subfigure}%
\begin{subfigure}{.33\textwidth}
  \includegraphics[scale=0.37]{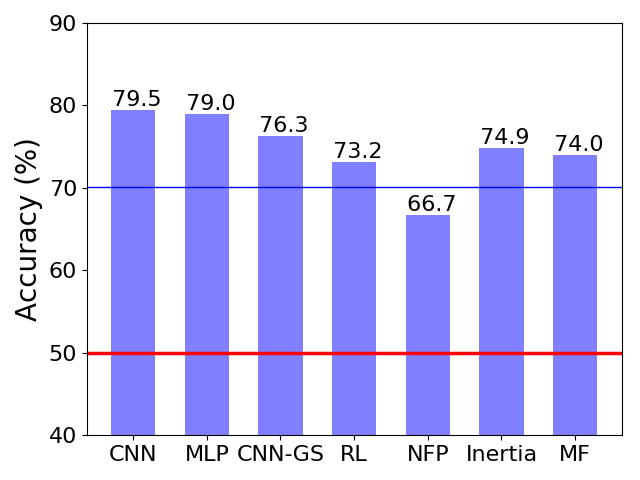}
  \caption{\label{fig:bar_dynamic_acc}}
\end{subfigure}
\begin{subfigure}{.33\textwidth}
  \includegraphics[scale=0.37]{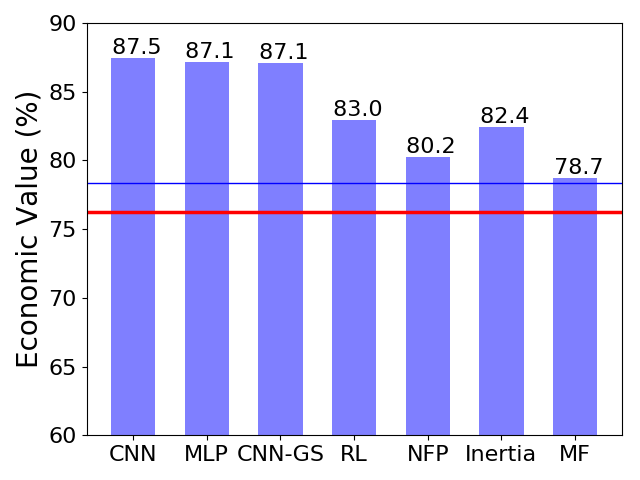}
  \caption{\label{fig:bar_dynamic_value}}
\end{subfigure}
\caption{
Comparison to dynamic models and benchmarks: (\ref{fig:bar_static_loss}) cross-entropy loss, (\ref{fig:bar_static_acc}) prediction accuracy, and (\ref{fig:bar_static_value}) economic value of the network models, the Reinforcement Learning (RL) and Normalized Fictitious Play (NFP) dynamic models, and the benchmarks of Inertia and the Most Frequent action in the previous history (MF). The blue horizontal line indicates the performance of the best static distribution benchmark, and the red line indicates the performance of the random benchmark (see Section \ref{dynamic} for more details). 
 \label{fig:bar_dynamic}}
\end{figure*}

\subsection{Comparison to Static Predictions} \label{static}

In order to successfully predict the dynamic step-by-step behavior, the networks 
must first be able to 
be more accurate 
than any static predictor that predicts a fixed distribution over the players' actions at all 
game rounds. We thus start by comparing the networks with stationary models and benchmarks. 

We compare with the following static models and benchmarks: the benchmark of the best possible static prediction, which is the empirical distribution of actions actually played by the players; the random benchmark, which predicts that players play at each step each of their actions uniformly at random; 
Nash Equilibrium (NE), which is the standard game theoretic equilibrium concept which assumes that each player perfectly best responds to the other players so as to maximize his payoff; 
and all four behavioral equilibrium models that were studied in \cite{SC2008}: 
(1) Quantal-Response Equilibrium (QRE), which extends Nash equilibrium by allowing the players to make mistakes \cite{qre}; 
(2) Payoff-Sampling equilibrium (PSE), in which players optimize against samples of their payoffs from each of their own pure strategies \cite{pse}; 
(3) Action-Sampling Equilibrium (ASE), in which players optimize against samples of strategies played by the other players \cite{SC2008}; 
and (4) Impulse-Balance Equilibrium (IBE), in which players respond to impulses of the difference between the payoff that could have been obtained and the payoff that has been received, such that for all players expected upward impulses are equal to expected downward impulses \cite{ibe}. 
\cite{SC2008} have shown that all of these four behavioral concepts predict human play better than the Nash equilibrium model.

Figures \ref{fig:bar_static_loss} and \ref{fig:bar_static_acc} present the loss and the accuracy of the 
two network types -- MLP and CNN 
-- alongside the two benchmarks and the minimal and maximal performance of the five equilibrium models we consider. 
As can be seen, 
both network types have significantly lower loss and higher accuracy levels than the best static distribution. 
Figure \ref{fig:bar_static_value} shows that this advantage is translated to a large difference also in the economic value: the network models achieve above 87\% of the optimal value, which is significantly higher than the 78.3\% achieved by the best static distribution. This is despite the fact that the networks were unaware of the utilities in the game during training or prediction. 
The CNN models perform better than the MLP models, with an advantage that is statistically significant according to each of the three measures at the 3\% level (paired-sample t-tests, $N=12$).
The results for the individual games show that the advantage of the neural networks was robust according to all metrics across all 12 games in our dataset (see the supplementary materials). 

The comparison between the five equilibrium models shows that in terms of prediction accuracy 
they perform similarly, all close to the best static distribution (Figure \ref{fig:bar_static_acc}). 
However, in terms of cross-entropy loss (Figure \ref{fig:bar_static_loss}), all equilibrium models had higher loss than the best static distribution (paired-sample t-tests, $N=12$, $p<0.05$ except for IBE for which $p<0.08$), and in terms of the economic value that they achieve, their performances vary between a value as low as the value achieved by the random benchmark and the higher value of the best static distribution. 
As could be expected, the worst performance is obtained by the Nash equilibrium model, which assumes rationality and is known to be less suitable for describing human behavior than behavioral equilibrium models that are based on behavioral tendencies.

\begin{figure*}[t] 
\begin{subfigure}{.34\linewidth} 
  \includegraphics[scale=0.39]{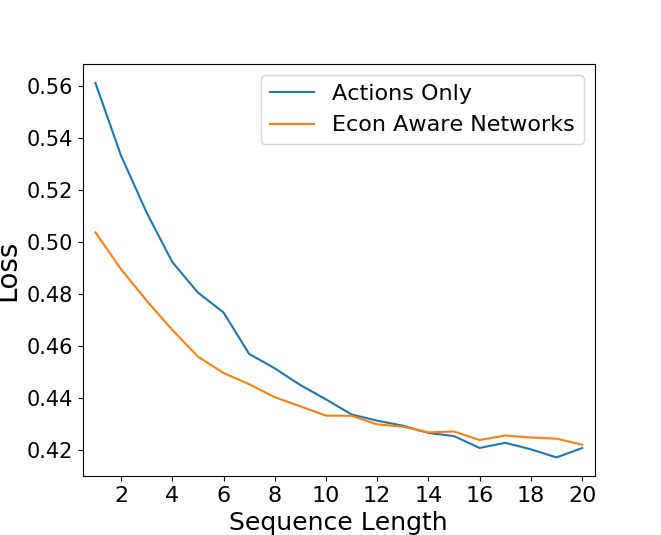} 
  \caption{\label{fig:seq_length_loss}}
\end{subfigure}%
\begin{subfigure}{.33\linewidth}
  \includegraphics[scale=0.39]{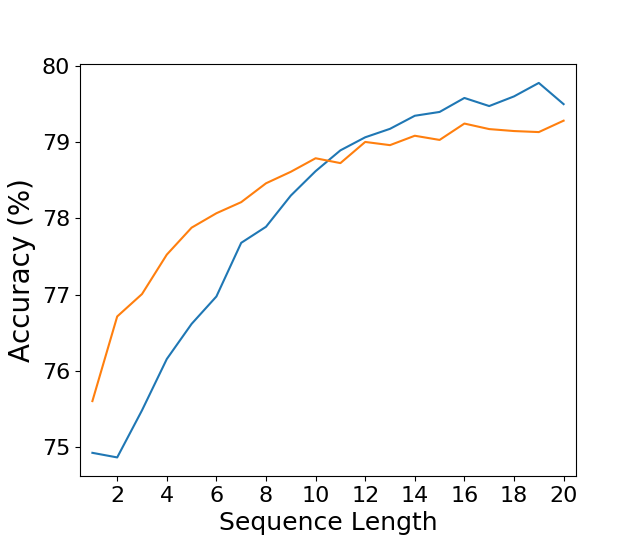}
  \caption{\label{fig:seq_length_acc}}
\end{subfigure}%
\begin{subfigure}{.33\linewidth}
  \includegraphics[scale=0.39]{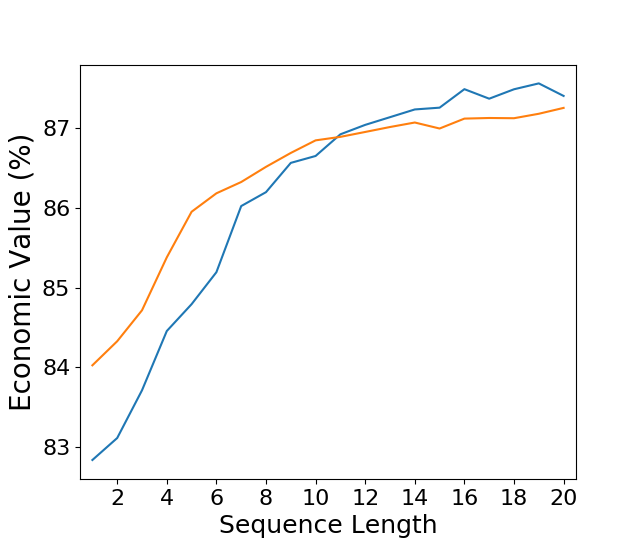}
  \caption{\label{fig:seq_length_value}}
\end{subfigure}
\caption{Econ-aware networks vs. networks that observe only the actions of the players: the CNN networks' performance as a function of the history sequence length given as input, 
in terms of (\ref{fig:seq_length_loss}) cross-entropy loss, (\ref{fig:seq_length_acc}) prediction accuracy, and (\ref{fig:seq_length_value}) the economic value obtained in percentages of the optimal value in the games (see Section \ref{econ} for more details). 
 \label{fig:seq_length}}
\end{figure*}

\subsection{Comparison to Dynamic Predictions} \label{dynamic}

The fact that the network models perform better than the best static distribution predictor indicates that they do capture some of the game dynamics and temporal correlations, and do not only learn stationary distributions. The real competition however is comparing them to other dynamic models which change their predictions according to changing inputs. 
The dynamic models and benchmarks we compare with the neural network models are 
the classic Reinforcement Learning (RL) model of \cite{rl} and the Normalized Fictitious Play (NFP) model of \cite{nfp} which outperformed all other fictitious play and reinforcement learning models in \cite{nfp2010}. Both models update their predicted distribution at every step according to the payoff results in the previous steps. 	
The two additional benchmarks are the Inertia rule that predicts that the player will maintain the same action that was played in the previous step, and the Most-Frequent (MF) heuristic, which is a generalization of the inertia rule that predicts that a player would play the action that he most frequently played in the previous $k$ steps. 
The parameters of RL, NFP and MF models were optimized with a grid search for every game according to the other 11 games.\footnote{We distinguish between the Inertia and MF benchmarks by restricting the optimization of MF to use $k>1$.}

We also compare our network models, which did not observe the test game at the training stage, to a Game-Specific CNN network (CNN-GS) that was trained only on sequences of the same game as the test game. In this case, for each game the networks were trained on 11 sessions in games 1-6, and on 5 session in games 7-12, and were tested on the remaining independent session. The performance in each game is then the average performance of the models over the sessions, and the overall performance of CNN-GS is the average of performances in all games (for each performance metric, weighted by the number of sessions). For game-specific networks there is a tradeoff between the specific information that the network observes about behavior in the specific game, and the reduced amount of data that is available for this game only.

Figures \ref{fig:bar_dynamic_loss} and \ref{fig:bar_dynamic_acc} present the loss and accuracy of the 
MLP and the CNN networks, 
compared with the dynamic models and benchmarks and with the game-specific networks. 
As can be seen, all the neural-network models outperform all the non-network models in terms of prediction loss and accuracy. 
The CNN models have the best performance, with an overall loss of 0.42 and accuracy of 79.5\%. 
Figure \ref{fig:bar_dynamic_value} shows the economic value obtained using the predictions of each model. We see that all the dynamic models and benchmarks yield higher economic value than static distribution predictors, and that again the neural network models 
outperform all the non-network models and benchmarks, 
with the best performance of 87.5\% of the optimal value that could have been obtained in the games achieved by the CNN models.
The advantage of the CNN models was statistically significant according to all three metrics (paired-sample t-tests, $N=12$, $p<0.03$ in comparison with MLP and $p<0.0001$ in all other comparisons except for the comparison with CNN-GS on the economic value for which $p=0.23$).\footnote{All differences seen in Figure \ref{fig:bar_dynamic} were statistically significant at least at the 3\% level (paired sample t-tests, $N=12$), except for the tests that compare the loss of NFP vs. the best-static distribution; the accuracy of CNN-GS vs. Inertia and RL vs. MF; and the economic value of CNN-GS vs. the two other network models, RL vs. Inertia, NFP vs. MF and MF vs. the best-static distribution.}

A comparison of the specialized network models (CNN-GS) to the domain-adapted network models (CNN and MLP) shows that the specialization-data tradeoff was not in favor of these networks. The CNN-GS achieved worse performance than the other networks according to cross-entropy loss and accuracy, 
but still outperformed all non-network models and benchmarks. 
When testing CNN-GS models only on the six 12-session games -- excluding the games with only 6 sessions, where the data tradeoff is more severe -- we see an improvement in the performance to a loss of 0.448, accuracy of 77.6\%, and economic value of 87.4\%.  

\section{Economic Features}  \label{econ}

The results in the previous section show that it is possible to predict future actions of individual players based solely on their observed history of actions, and that neural networks can successfully generalize and perform the prediction on previously unseen games at almost 80\% accuracy level, without using any information about the structure of the new game.  
Now we ask whether adding economic information improves the quality of the predictions. 
For this we train econ-aware models, that in addition to the input of the actions that were played  get as input the obtained and forgone payoffs of both players in a given interval of game history, and the payoff matrix of the game. 
We compare the performance of these models to the action-only aware CNN models when given as input history sequences of different lengths $k$.

Figure \ref{fig:seq_length} shows the performance of the econ-aware models and of the action-only aware models as a function of the length of the history sequence input, in terms of cross-entropy loss, accuracy and economic value. 
The orange lines indicate the performance of the econ-aware models and the blue lines are of the action-only aware models, averaged over the models of the 12 games. 
First we see that both models improve their performance when given inputs of a longer history sequence. This improvement saturates around a sequence length of 20 game periods.\footnote{We also tested sequences of up to 100 periods and these did not improve the results.}
For short sequence lengths, as could be expected, the ability of the action-only aware models is impaired, and for an input of length 1 these networks do not perform much better than the inertia rule (see Figure \ref{fig:bar_dynamic}). For these short sequence length inputs, additional economic information significantly improves the ability of the networks to predict the next steps according to all three metrics. 
However, when looking at models trained on longer sequences, the additional economic information no longer improves the prediction results. 
At sequence lengths of above 11 game rounds, the action-only aware models succeed in the prediction task better than the econ-aware models according to all metrics: loss, accuracy and economic value. 

We see that, at least in our setting, a history of the past 20 steps is sufficient for predicting future actions of players. Although at first glance this seems like a rather short period of time to encode all the considerations that our human players may have, note that actually a sequence of length 20 with two players and two actions allows for $4^{20}$ (about a trillion) possible sequences that are all allowed by the game rules, which is a very rich phase space. The fact that conditional prediction is possible at all in this vast space of configurations indicates that there are indeed (possibly complicated) regularities in the way human players play, that are general across different games. 

\section{Conclusion}  \label{conclusion}

In this study we addressed the challenge of predicting the actions of individual human players in repeated games, and generalizing this prediction to previously unseen games. 
We found that 
neural networks are able to perform this task well, and that even when they are given only the players' empirical actions as input they 
outperform 
behavioral models that explicitly take into account the economic considerations of the players and aim to model their decision-making processes. 
The advantage of the neural networks was with a significant gap in terms of all three metrics that we considered: cross-entropy loss, prediction accuracy, and economic value. 

We further showed that when the networks are trained on short sequences of history of play, economic information about the utilities in the game improves their prediction performance, as could be expected. 
However, interestingly, when trained on long enough sequences, this information is no longer required, and in fact the networks that are based only on the players' actions perform better than the econ-aware networks according to all three metrics. 
Of course, it is unlikely that the players ignore the economic information of the game, but our results indicate that the sequences of joint action profiles encode all the information that is useful for the prediction of the players' next actions.
This result may be particularly useful in other domains, such as predicting human players' decisions in online auctions, where there are large amounts of data of the players' previous actions, but their utilities are their private information and are not available in the data. 
This observation is conceptually inline with recent works in econometrics showing that inference of players' private parameters should be based on the action distribution rather than on specific equilibrium calculations \cite{quantalregret}.
It is interesting to study further this connection between the input to neural networks and their ability to predict human players' decision making, and test if similar results hold in other contexts 
as well and whether it is possible to improve the prediction performance by using other network architectures. 

Finally, as in other domains where neural networks are successfully utilized for prediction, they have the disadvantage of working as a black box, in contrast to behavioral models that explicitly describe the decision-making processes of human players; yet, the success of networks in the prediction task, and the gap we have shown above the performance of the behavioral models, may provide insights regarding the degree of regularities that exist in data of human interactions -- regularities that may then be captured by improved  behavioral models.

\section*{Acknowledgments}

The authors thank Noam Nisan, David Parkes and Amit Danieli for useful discussions. 
This project has received funding from the European Research Council (ERC) under the European Union's Horizon 2020 research and innovation programme (grant agreement No 740282). 
Gali Noti is supported by the Adams Fellowship Program of the Israel Academy of Sciences and Humanities.

\bibliographystyle{named}
\bibliography{networks-bib}

\end{document}